\documentclass[12pt,thmsa]{article}
\usepackage{amssymb}

\usepackage{sw20lart}

\input{tcilatex}
\begin{document}

\author{Emilio Santos \and Departamento de F\'{i}sica. Universidad de Cantabria.
Santander. Spain}
\title{Constraints on fourth order generalized f(R) gravity}
\date{August, 4, 2011 }
\maketitle

\begin{abstract}
A fourth order generalized f(R) gravity theory (FOG) is considered with the
Einstein-Hilbert action $R+aR^{2}+bR_{\mu \nu }R^{\mu \nu },$ $R_{\mu \nu }$
being Ricci\'{}s tensor and R the curvature scalar. The field equations are
applied to spherical bodies where Newtonian gravity is a good approximation.
The result is that for $0\leq a\sim -b<<R^{2}$, $R$ being the body radius,
the gravitational field outside the body contains two Yukawas, one
attractive and the other one repulsive, in addition to the Newtonian term.
For $a\sim -b>>R^{2}$ the gravitational field near the body is zero but at
distances greater than $\sqrt{a}\sim \sqrt{-b}$ the field is practically
Newtonian. From the comparison with laboratory experiments I conclude that $%
\sqrt{a}$ and $\sqrt{-b}$ should be smaller than a few millimeters, which
excludes any relevant effect of FOG on stars, galaxies or cosmology.
\end{abstract}

\section{ Fourth order generalized f(R) gravity and quantum vacuum}

In recent years a great effort has been devoted to extended gravity theories
which modify general relativity. The main motivation was the search for
physical explanations to the observed accelerated expansion of the universe
and other astrophysical observations, like the flat rotation curves in
galaxies \cite{Sahni}. It is plausible to derive the extended theory from a
generalized Einstein-Hilbert action 
\begin{equation}
S=\frac{1}{2k}\int d^{4}x\sqrt{-g}\left( R+F\right) +S_{mat},  \label{0a}
\end{equation}
where $F$ should be a function of the scalars which may be obtained by
combining the Riemann tensor, $R_{\mu \nu \lambda \sigma },$ and its
derivatives, with the metric tensor, $g_{\mu \nu }$. The theory derived from
the particular choice $F(R),$ where $R$ is the Ricci scalar, has been
extensively explored under the name of \textit{f(R)-gravity}\cite{Faraoni}.
But it is possible to consider more general forms of $F$, e. g. depending on
the Ricci tensor in addition to the curvature scalar. The most simple
extension of general relativity seems to be \textit{fourth order gravity (FOG%
}), where the functional $F$ has the form 
\begin{equation}
F=aR^{2}+bR_{\mu \nu }R^{\mu \nu },  \label{a1}
\end{equation}
where $G_{\mu \nu }$ is the Einstein tensor, $R_{\mu \nu }$ the Ricci
tensor, $R$ the curvature scalar and $k$ is $8\pi $ times Newton\'{}s
constant, $a$ and $b$ being two constant parameters with dimensions of
length squared\cite{Santosa}, \cite{Stabile}.

Another approach to FOG comes from the assumption that the energy of the
quantum vacuum is not zero in curved spacetime\cite{Birrell}, \cite{Wald}.
This fact may be taken into account by adding a new stress-energy tensor, $%
T_{\mu \nu }^{vac},$ to the matter one, $T_{\mu \nu }^{mat},$ so that
Einstein\'{}s equation becomes 
\begin{equation}
G_{\mu \nu }\equiv R_{\mu \nu }-\frac{1}{2}g_{\mu \nu }R=k\left( T_{\mu \nu
}^{mat}+T_{\mu \nu }^{vac}\right) ,  \label{E}
\end{equation}
Eq.$\left( \ref{E}\right) $ may be derived from the action $\left( \ref{0a}%
\right) ,$ the tensor $T_{\mu \nu }^{vac}$ coming from the functional $F$.
In not too strong gravitational fields a plausible form of $F$ is given by
eq.$\left( \ref{a1}\right) $\cite{Santosa}. In summary, we see that FOG may
be seen as either an extension of general relativity or a quantum vacuum
effect. In the latter case the tensor $T_{\mu \nu }^{vac}$ appears on the
right side of eq.$\left( \ref{E}\right) ,$ in the former it would appear on
the left. But in both cases it gives rise to the same physical theory. In
this paper I will use a language corresponding to the quantum vacuum
assumption.

The modifications derived from $F$ should be small in weak gravitational
fields, where GR is valid, but might be relevant in more strong fields like
those existing in compact stars or the early universe. In order to study
that possibility we should find the range or the parameters $a$ and $b$
compatible with known data, which is the purpose of this paper.

\section{Field equations}

The tensor field equation derived from the functional $\left( \ref{a1}%
\right) $ may be taken from the literature\cite{Stabile}. I shall write it
in terms of the Einstein tensor, $G_{\mu \nu }$, rather than the Ricci
tensor, $R_{\mu \nu },$ and in a form that looks like the standard Einstein
equation of general relativity eq.$\left( \ref{E}\right) $. That is 
\[
G_{\mu \nu }=kT_{\mu \nu },\;T_{\mu \nu }=T_{\mu \nu }^{mat}+T_{\mu \nu
}^{vac}, 
\]

\begin{eqnarray}
&&kT_{\mu \nu }^{vac}\equiv -(2a+b)\left[ \nabla _{\mu }\nabla _{\nu
}G-g_{\mu \nu }\Box G\right] -2\left( a+b\right) \left[ -GG_{\mu \nu }+\frac{%
1}{4}g_{\mu \nu }G^{2}\right]  \nonumber \\
&&-b\left[ 2G_{\mu }^{\sigma }G_{\sigma \nu }-\frac{1}{2}g_{\mu \nu
}G_{\lambda \sigma }G^{\lambda \sigma }-\nabla _{\sigma }\nabla _{\nu
}G_{\mu }^{\sigma }-\nabla _{\sigma }\nabla _{\mu }G_{\nu }^{\sigma }+\Box
G_{\mu \nu }\right] ,  \label{1}
\end{eqnarray}
with obvious notation.

We are interested in static problems of spherical symmetry and will use the
standard metric

\begin{equation}
ds^{2}=-\exp \left( \beta \left( r\right) \right) dt^{2}+\exp \left( \alpha
\left( r\right) \right) dr^{2}+r^{2}d\Omega ^{2}.  \label{metric}
\end{equation}
Thus $G_{\mu \nu }\left( r\right) $ and $T_{\mu \nu }^{mat}\left( r\right) $
have 3 independent components each, so that including $\alpha \left(
r\right) $ and $\beta \left( r\right) $ there are 8 unknown functions of $r$%
. On the other hand there are 8 equations, namely 3 eqs.$\left( \ref{1}%
\right) ,$ 3 more equations giving the independent components of $G_{\mu \nu
}$ in terms of $\alpha $ and $\beta $ and 2 equations of state relating the
3 independent components of $T_{\mu \nu }^{mat}.$ I shall assume local
isotropy for matter, so that one of the latter will be the equality $%
T_{11}^{mat}=T_{22}^{mat}$ ($=T_{33}^{mat}$ in spherical symmetry.) In
principle the remaining 7 coupled non-linear equations may be solved exactly
by numerical methods.

Before proceeding, a note about the signs convention is in order. As is well
known different authors use different signs in the definition of the
relevant quantities. Here I shall make a choice which essentially agrees
with the one of Ref.\cite{Faraoni}. It may be summarized as follows

\begin{equation}
g_{00}=-\exp \beta ,G_{\mu \nu }=R_{\mu \nu }-\frac{1}{2}g_{\mu \nu
}R=kT_{\mu \nu },T_{0}^{0}=-\rho .  \label{signs}
\end{equation}
After that I shall write the three independent components of eq.$\left( \ref
{1}\right) $ using the notation 
\begin{eqnarray}
T_{0}^{0} &=&-\rho ,T_{1}^{1}=p,T_{2}^{2}=q,T_{\mu }^{\mu }=T=p+2q-\rho , 
\nonumber \\
\left( T_{mat}\right) _{0}^{0} &=&-\rho _{mat},\left( T_{mat}\right)
_{1}^{1}=\left( T_{mat}\right) _{2}^{2}=\left( T_{mat}\right)
_{3}^{3}=p_{mat}.  \label{ropq}
\end{eqnarray}
In the following I will name $\rho ,p$ and $q$ the \textit{total} density,
radial pressure and transverse pressure respectively, whilst $\rho _{mat}$
and $p_{mat}$ will be named \textit{matter} density and pressure
respectively (remember that we assume local isotropy for matter, that is
equality of radial and transverse matter pressures.) The differences $\rho
-\rho _{mat},p-p_{mat}$ and $q-p_{mat}$ will be named \textit{vacuum}
density, radial pressure and transverse pressure, respectively.

After some algebra I get for the components of the tensor eq.$\left( \ref{1}%
\right) $ as follows

\begin{eqnarray}
-\rho _{mat} &=&-\rho +(2a+b)e^{-\alpha }\left[ -\frac{d^{2}T}{dr^{2}}%
-\left( \frac{2}{r}-\frac{1}{2}\alpha ^{\prime }\right) \frac{dT}{dr}\right]
\nonumber \\
&&+(a+b)k(\frac{1}{2}T^{2}+2T\rho )+b\left[ -\Delta \rho +2k\rho ^{2}-\frac{1%
}{2}k\left[ \rho ^{2}+p^{2}+2q^{2}\right] \right]  \nonumber \\
&&+b\exp (-\alpha )\left[ -\frac{2\beta ^{\prime }}{r}\left( \rho +q\right)
+\left( \frac{1}{2}\alpha ^{\prime }\beta ^{\prime }-\beta ^{\prime \prime
}\right) \left( \rho +p\right) \right] ,\smallskip  \label{rom}
\end{eqnarray}
\begin{eqnarray}
p_{mat} &=&p-(2a+b)e^{-\alpha }\left( \frac{2}{r}+\frac{1}{2}\beta ^{\prime
}\right) \frac{dT}{dr}+(a+b)k(\frac{1}{2}T^{2}-2Tp)  \nonumber \\
&&+b\left[ \Delta p+2kp^{2}-\frac{1}{2}k\left[ \rho ^{2}+p^{2}+2q^{2}\right]
\right]  \nonumber \\
&&+b\exp (-\alpha )\left[ \left( \frac{2\alpha ^{\prime }}{r}+\frac{4}{r^{2}}%
\right) \left( q-p\right) +\left( -\frac{1}{2}\alpha ^{\prime }\beta
^{\prime }+\beta ^{\prime \prime }\right) (\rho +p)\right] ,  \label{pm}
\end{eqnarray}
\begin{eqnarray}
p_{mat} &=&q-(2a+b)e^{-\alpha }\left[ \frac{d^{2}T}{dr^{2}}+\left( \frac{1}{r%
}+\frac{1}{2}\beta ^{\prime }-\frac{1}{2}\alpha ^{\prime }\right) \frac{dT}{%
dr}\right]  \nonumber \\
&&+(a+b)k(\frac{1}{2}T^{2}-2Tq)+b\left[ \Delta q+2kq^{2}-\frac{1}{2}k\left[
\rho ^{2}+p^{2}+2q^{2}\right] \right]  \nonumber \\
&&+b\exp (-\alpha )\left[ \left( -\frac{\alpha ^{\prime }}{r}-\frac{2}{r^{2}}%
\right) \left( q-p\right) +\frac{\beta ^{\prime }}{r}(\rho +q)\right] .
\label{qm}
\end{eqnarray}
Addition of these 3 equations gives the trace equation, that is 
\begin{equation}
T_{mat}\equiv 3p_{mat}-\rho _{mat}=T-\left( 6a+2b\right) \Delta T,
\label{trace}
\end{equation}
where $\Delta $ is the Laplacean operator in curved space-time, that is 
\begin{equation}
\Delta \equiv \exp (-\alpha )\left[ \frac{d^{2}}{dr^{2}}+\left( \frac{2}{r}+%
\frac{1}{2}\beta ^{\prime }-\frac{1}{2}\alpha ^{\prime }\right) \frac{d}{dr}%
\right] .  \label{laplace}
\end{equation}
The quantities G$_{\mu }^{\nu }$ are related to the metric coefficients $%
\alpha $ and $\beta $ and their derivatives, hence to $\rho ,p$ and $q,$
that is 
\begin{eqnarray}
\exp (-\alpha ) &=&1-\frac{2m}{r},\frac{\alpha ^{\prime }}{2}=\frac{m-4\pi
\rho r^{3}}{r^{2}-2mr},\beta ^{\prime }=2\frac{m+4\pi r^{3}p}{r^{2}-2mr}, 
\nonumber \\
\beta ^{\prime \prime } &=&\frac{8\pi r^{2}\left( r\rho +rp+3p^{\prime
}\right) }{r^{2}-2mr}-\frac{4\left( m+4\pi r^{3}p\right) \left( r-m-4\pi
r^{3}\rho \right) }{\left( r^{2}-2mr\right) ^{2}},  \label{alfabeta}
\end{eqnarray}
where I have used units $k=8\pi ,c=1$ and the radial derivative of $\alpha
\left( \beta ^{\prime }\right) $ is labelled $\alpha ^{\prime }\left( \beta
^{\prime \prime }\right) $. The mass parameter $m$ is defined by 
\begin{equation}
m=\int_{0}^{r}4\pi x^{2}\rho (x)dx.  \label{mass}
\end{equation}
The condition that Einstein tensor, $G_{\mu \nu },$ is divergence free leads
to the hydrostatic equilibrium equation, that is 
\begin{equation}
\frac{dp}{dr}=\frac{2(q-p)}{r}-\frac{1}{2}\beta ^{\prime }\left( \rho
+p\right) .  \label{OV}
\end{equation}

\section{Terrestrial constraints on fourth order gravity (FOG)}

The theory derived from $F=0$ in the action $\left( \ref{0a}\right) ,$ that
is general relativity, is known to give good agreement with observations for
a wide range of intensities of the gravitational field (that is curvature of
spacetime.) As a consequence the corrections due to finite, nonzero, values
of the parameters $a$ and $b$ should be below the uncertainties in the data
in that domain. In a previous paper \cite{Santosa} I considered the problem,
but there are additional constraints not taken into account there, which
makes necessary a more detailed study.

In order to study the constraints on $a$ and $b$ derived from terrestrial
and solar system observations I shall start solving eqs.$\left( \ref{rom}%
\right) $ to $\left( \ref{OV}\right) $ for spherical bodies - like the Sun,
the Earth or a laboratory sphere of metal - where Newtonian gravity is a
fairly good approximation. Corrections to Newtonian gravity coming from FOG
have been obtained via the Newtonian approximation of the field eqs.$\left( 
\ref{rom}\right) $ to $\left( \ref{qm}\right) $\cite{Stabile}, \cite
{Odintsov}. Here I shall calculate the gravitational field by solving
directly the field equations, with appropriate approximations.

For the Earth the parameters $\alpha $ and $\beta $ of the metric $\left( 
\ref{metric }\right) $ are very close to unity and terms like $GG_{\mu
}^{\nu }$ or $g_{\mu }^{\nu }G^{2}$ are smaller than the main term, $G_{\mu
}^{\nu },$ by about 
\[
ak\rho _{mat}/c^{2}\sim bk\rho _{mat}/c^{2}\lesssim 10^{-26}, 
\]
for $a\sim b\lesssim 1$ m$^{2}$. In addition the matter pressure, $p_{mat},$
is negligible in comparison with matter density, $\rho _{mat}.$ As a
consequence eqs.$\left( \ref{rom}\right) $ to $\left( \ref{qm}\right) $ may
be approximated by the following 
\begin{equation}
b\left[ \frac{d^{2}\rho }{dr^{2}}+\frac{2}{r}\frac{d\rho }{dr}\right]
+(2a+b)\left[ \frac{d^{2}T}{dr^{2}}+\frac{2}{r}\frac{dT}{dr}\right] +\rho
=\rho _{mat},  \label{rom1}
\end{equation}
\begin{equation}
-b\left[ \frac{d^{2}p}{dr^{2}}+\frac{2}{r}\frac{dp}{dr}+\frac{4}{r^{2}}%
\left( q-p\right) \right] +(2a+b)\frac{2}{r}\frac{dT}{dr}-p=0,  \label{pm1}
\end{equation}
\begin{equation}
-b\left[ \frac{d^{2}q}{dr^{2}}+\frac{2}{r}\frac{dq}{dr}-\frac{2}{r^{2}}%
\left( q-p\right) \right] +(2a+b)\left[ \frac{d^{2}T}{dr^{2}}+\frac{1}{r}%
\frac{dT}{dr}\right] -q=0.  \label{qm1}
\end{equation}
Eqs.$\left( \ref{rom1}\right) $ to $\left( \ref{qm1}\right) $ toghether with
eqs.$\left( \ref{alfabeta}\right) $ to $\left( \ref{OV}\right) ,$ plus the
equation of state, are a system of coupled differential equations whose
solution is involved. However a great simplification is possible if we
assume that the corrections due to finite values of the parameters $a$ and $%
b $ would modify but slightly the function $\rho _{mat}\left( r\right) $
with respect to the results obtained from a Newtonian treatment. Thus we may
take $\rho _{mat}\left( r\right) $ as given, which decouples eqs.$\left( \ref
{rom1}\right) $ to $\left( \ref{qm1}\right) $ from the remaining ones. Still
the three eqs.$\left( \ref{rom1}\right) $ to $\left( \ref{qm1}\right) $ are
coupled amongst themselves, but from them it is possible to get two
decoupled ones. In fact if I add eq.$\left( \ref{rom1}\right) ,$ eq.$\left( 
\ref{pm1}\right) $ plus twice eq.$\left( \ref{qm1}\right) $ I get the trace
eq.$\left( \ref{trace}\right) $, which may be rewritten 
\begin{equation}
\left( 6a+2b\right) \left[ \frac{d^{2}T}{dr^{2}}+\frac{2}{r}\frac{dT}{dr}%
\right] -T=-T_{mat}\simeq \rho _{mat}.  \label{trace1}
\end{equation}
Subtracting this minus three times eq.$\left( \ref{rom1}\right) $ I obtain 
\begin{equation}
b\left[ \frac{d^{2}(T+3\rho )}{dr^{2}}+\frac{2}{r}\frac{d(T+3\rho )}{dr}%
\right] +T+3\rho =2\rho _{mat}.  \label{Tro}
\end{equation}

Now the general solution of the trace eq.$\left( \ref{trace1}\right) $ is
trivial and I get 
\begin{eqnarray}
T\left( r\right) &=&\frac{1}{\sqrt{6a+2b}r}\int_{0}^{r}\sinh \left( \frac{r-z%
}{\sqrt{6a+2b}}\right) z\rho _{mat}\left( z\right) dz  \nonumber \\
&&+\frac{A}{r}\exp \left( \frac{r}{\sqrt{6a+2b}}\right) +\frac{B}{r}\exp
\left( -\frac{r}{\sqrt{6a+2b}}\right) .  \label{T}
\end{eqnarray}
The integration constants $A$ and $B$ may be got from the boundary
conditions, that is the function $\rho \left( r\right) $ should be finite at
the origin and go to zero at infinity, which gives 
\begin{equation}
A=-B=-\frac{1}{2\sqrt{6a+2b}}\int_{0}^{R}\exp \left( -\frac{z}{\sqrt{6a+2b}}%
\right) z\rho _{mat}\left( z\right) dz,  \label{AB}
\end{equation}
where $R$ is the radius of the body, that is, the radius of the matter
distribution. Note that a part of the total mass of the body lies in the
region without matter, associated to the vacuum density. Eq.$\left( \ref{Tro}%
\right) $ may be solved by a method similar to the one used for eq.$\left( 
\ref{trace1}\right) ,$ which gives the function $\rho \left( r\right) .$
Once we know the function $T(r)$ the solutions of eqs.$\left( \ref{pm1}%
\right) $ and $\left( \ref{qm1}\right) $ are straightforward.

I am interested in the functions $T\left( r\right) ,\rho \left( r\right)
,p(r)$ and $q(r)$ outside the body (i.e. $r>R,$ where $\rho _{mat}=0)$ and
for the particular case where $\rho _{mat}$ is a constant inside the body (
which is a good approximation for the Earth or a metallic sphere). Thus the $%
z$ integrals in eqs.$\left( \ref{T}\right) $ and $\left( \ref{AB}\right) $
may be performed analytically, and similarly in the solution of eq.$\left( 
\ref{Tro}\right) .$ In order to simplify the notation I shall introduce the
following dimensionless variables 
\begin{equation}
x\equiv \frac{r}{\sqrt{6a+2b}},X\equiv \frac{R}{\sqrt{6a+2b}},y\equiv \frac{r%
}{\sqrt{-b}},Y\equiv \frac{R}{\sqrt{-b}},  \label{xy}
\end{equation}
and the mass parameters 
\begin{equation}
M_{x}\equiv \frac{3M}{2X^{3}}\left[ X-1+(X+1)\exp \left( -2X\right) \right] ,%
\text{ }M_{y}\equiv \frac{3M}{2Y^{3}}\left[ Y-1+(Y+1)\exp \left( -2Y\right)
\right] .  \label{M}
\end{equation}
Thus I get for $r>R$%
\begin{equation}
T\left( r\right) =-\frac{M_{x}}{4\pi \left( 6a+2b\right) r}\exp \left(
X-x\right) ,  \label{out}
\end{equation}
\begin{equation}
\rho (r)=\frac{1}{12\pi r}\left[ \left( 6a+2b\right) ^{-1}M_{x}\exp \left(
X-x\right) +2\left| b\right| ^{-1}M_{y}\exp \left( Y-y\right) \right] .
\label{roout}
\end{equation}
Hence it follows 
\begin{equation}
p+2q=T+\rho =\frac{1}{6\pi r}\left[ \left| b\right| ^{-1}M_{y}\exp \left(
Y-y\right) -\left( 6a+2b\right) ^{-1}M_{x}\exp \left( X-x\right) \right] .
\label{p2q}
\end{equation}
In order to get $p$ and $q$ separately we shall solve the equation resulting
from the subtraction of eq.$\left( \ref{qm1}\right) $ minus eq.$\left( \ref
{pm1}\right) $ in the region $r>R,$ that is 
\begin{eqnarray}
\left| b\right| \left[ \frac{d^{2}h}{dr^{2}}+\frac{2}{r}\frac{dh}{dr}-\frac{%
6h}{r^{2}}\right] -h &=&(2a+b)(\frac{1}{r}\frac{dT}{dr}-\frac{d^{2}T}{dr^{2}}%
)  \label{h} \\
&=&\frac{\left( 2a+b\right) M_{x}}{4\pi (6a+2b)^{5/2}}\left( 1+\frac{3}{x}+%
\frac{3}{x^{2}}\right) \exp \left( X-x\right) ,  \nonumber
\end{eqnarray}
where eq.$\left( \ref{out}\right) $ has been taken into account. Now I use
the \textit{ansatz} 
\[
h=A\left( 1+\frac{3}{x}+\frac{3}{x^{2}}\right) \exp \left( X-x\right)
+B\left( 1+\frac{3}{y}+\frac{3}{y^{2}}\right) \exp \left( Y-y\right) , 
\]
which inserted in the left side of eq.$\left( \ref{h}\right) $ gives 
\[
A(\frac{\left| b\right| }{6a+2b}-1)\left( 1+\frac{3}{x}+\frac{3}{x^{2}}%
\right) \exp \left( X-x\right) . 
\]
This leads to 
\[
A(\frac{\left| b\right| }{6a+2b}-1)=\frac{\left( 2a+b\right) M_{x}}{6a+2b}%
\Rightarrow A=-\frac{M_{x}}{12\pi }, 
\]
whence I get 
\[
h=-\frac{M_{x}}{12\pi }\left( 1+\frac{3}{x}+\frac{3}{x^{2}}\right) \exp
\left( X-x\right) +B\left( 1+\frac{3}{y}+\frac{3}{y^{2}}\right) \exp \left(
Y-y\right) . 
\]
Combining this with eq.$\left( \ref{p2q}\right) $ I obtain 
\begin{eqnarray*}
p &=&\frac{1}{6\pi r}\left[ \left| b\right| ^{-1}M_{y}\exp \left( Y-y\right)
-\left( 6a+2b\right) ^{-1}M_{x}\exp \left( X-x\right) \right] \\
&&+\frac{M_{x}}{6\pi }\left( 1+\frac{3}{x}+\frac{3}{x^{2}}\right) \exp
\left( X-x\right) -2B\left( 1+\frac{3}{y}+\frac{3}{y^{2}}\right) \exp \left(
Y-y\right) .
\end{eqnarray*}

The integration constant $B$ is fixed by the condition that eq.$\left( \ref
{OV}\right) $ is fulfilled with the approximation of neglecting the last
term. That is 
\[
\frac{dp}{dr}=\frac{2(q-p)}{r}\Longleftrightarrow \frac{d}{dr}\left(
r^{3}p\right) =r^{2}\left( 2q+p\right) . 
\]
Finally this leads to

\begin{equation}
p\left( r\right) =\frac{M_{x}}{6\pi r^{3}}\left[ x+1\right] \exp \left(
X-x\right) -\frac{M_{y}}{6\pi r^{3}}\left[ y+1\right] \exp \left( Y-y\right)
.  \label{pout}
\end{equation}
All results obtained up to now for the solution of the differential eqs.$%
\left( \ref{rom1}\right) $ to $\left( \ref{Tro}\right) $ have assumed that
both $6a+2b$ and $-b$ are positive. If one or both quantities were negative
we would obtain sinus or cosinus functions rather than exponentials, which
would clearly violate empirical facts. Thus the parameters are constrained
to the range 
\begin{equation}
0\leq -b\leq 3a.  \label{range}
\end{equation}
The remaining bounds are derived in the following.

The vacuum density eq.$\left( \ref{roout}\right) $ gives rise to a mass
distribution which I calculate as follows. Firstly I show that the total
mass of the body is the same as the mass of matter, that is the total vacuum
mass is zero. This follows trivially if we integrate the two sides of eq.$%
\left( \ref{rom1}\right) $ after multiplication times the volume element. In
fact we obtain 
\[
\int_{0}^{\infty }\nabla ^{2}\rho 4\pi r^{2}dr=4\pi \int_{0}^{\infty }r\frac{%
d^{2}}{dr^{2}}\left( r\rho \right) dr=-4\pi \int_{0}^{\infty }\frac{d}{dr}%
\left( r\rho \right) dr=0, 
\]
where I have performed an integration by parts and taken into account that $%
r\rho \rightarrow 0$ for $r\rightarrow \infty $. A similar result holds true
in the integral of $\nabla ^{2}T.$ Thus the mass asociated to the total
density $\rho $ equals the mass asociated to the matter density $\rho
_{mat}. $ In our language we may say that \textit{the total mass of the
quantum vacuum (}associated to\textit{\ }$T_{\mu \nu }^{vac},$ see eq.$%
\left( \ref{1}\right) )$ \textit{is zero. }Nevertheless the vacuum density
changes the spatial distribution of the total mass. In fact, we may obtain
the (vacuum) mass distribution outside the body by integrating eq.$\left( 
\ref{roout}\right) $ between $R$ and $r$. I get 
\begin{eqnarray*}
M_{ext}\left( r\right) &=&4\pi \int_{R}^{r}r^{2}\rho dr \\
&=&\frac{M_{x}}{3}\left[ X+1-\left( x+1\right) \exp \left( X-x\right)
\right] +\frac{2M_{y}}{3}\left[ Y+1-\left( y+1\right) \exp \left( Y-y\right)
\right] .
\end{eqnarray*}
Hence it follows that the total mass enclosed in a sphere of radius $r$ is
(see eqs.$\left( \ref{xy}\right) $) 
\begin{equation}
M\left( r\right) =M-\frac{M_{x}}{3}\left( x+1\right) \exp \left( X-x\right) -%
\frac{2M_{y}}{3}\left( y+1\right) \exp \left( Y-y\right) ,  \label{masa}
\end{equation}
where the condition $M\left( \infty \right) =M$ has been taken into account.
Remember that $M\left( \infty \right) -$ $M$ corresponds to the \textit{%
vacuum} mass, which is zero as shown above. The effect is dramatic for a
spherical body with a radius $R<<$ $\sqrt{a},\sqrt{\left| b\right| }.$ In
this case $X,Y<<1$, $M_{x}\simeq M_{y}\simeq M$ which implies $M\left(
r\right) \simeq 0,$ that is \textit{the total mass in the interior of the
body is zero. That is the vacuum mass (negative) cancels the matter mass
(positive) in the interior of the body.}

Now we may calculate the gravitational field, $g$, near the Earth surface,
which we should identify with $-1/2$ times the quantity $\beta ^{\prime }$
defined in $\left( \ref{alfabeta}\right) ,$ where we may neglect $m<<r$ in
the denominator. Thus from eqs.$\left( \ref{roout}\right) $ and $\left( \ref
{pout}\right) $ I get, to lowest order in $\sqrt{6a+2b}$ and $\sqrt{\left|
b\right| },$

\begin{eqnarray}
g &=&-\frac{G}{r^{2}}\left[ M(r)+4\pi r^{3}p(r)\right]  \nonumber \\
&\simeq &-\frac{GM}{r^{2}}-\frac{GM_{x}}{3r^{2}}\left( x+1\right) \exp
\left( X-x\right) +\frac{4GM_{y}}{3r^{2}}\left( y+1\right) \exp \left(
Y-y\right) ,  \label{g}
\end{eqnarray}
where $G$ is Newton constant. We see that when $\sqrt{6a+2b},\sqrt{\left|
b\right| }<<R$ eq.$\left( \ref{g}\right) $ corresponds to Newtonian gravity
plus a small correction consisting of two Yukawa-type terms, one of them
attractive, the other one repulsive. Although these results have been
derived for bodies with constant density they are valid for any celestial
body because the parameters $M_{x}$ and $M_{y}$ depend only on a small
region near the body\'{}s surface where the density is effectively a
constant.

In sharp contrast for a small sphere where $\sqrt{6a+2b},\sqrt{\left|
b\right| }>>R$ and $M_{x}\simeq M_{y}\simeq M$ (see eqs.$\left( \ref{M}%
\right) )$ we obtain 
\begin{eqnarray*}
\left| g\right| &\simeq &\frac{GM}{r^{2}}\left| 1+\frac{1}{3}\exp \left(
X-x\right) -\frac{4}{3}\exp \left( Y-y\right) \right| \\
&\simeq &\frac{GM}{3r^{2}}\left( r-R\right) \left| \frac{1}{\sqrt{6a+2b}}-%
\frac{4}{\sqrt{\left| b\right| }}\right| <<\frac{GM}{r^{2}}
\end{eqnarray*}
for $r$ not much larger than both $\sqrt{6a+2b}$ and $\sqrt{\left| b\right| }
$. Now performed laboratory experiments have shown that the field $g$ agrees
fairly well with Newtonian predictions for bodies greater than a few
millimeters\cite{Fischbach}. Thus our calculation shows that these
experiments exclude values of the parameters $a$ and $\left| b\right| $
greater than a fraction of squared centimeter.

\section{Conclusions}

I conclude that $\sqrt{a}$ and $\sqrt{-b}$ should be smaller than one
centimeter, which probably excludes any relevant effect of FOG on stars,
galaxies or cosmology.


\begin{thebibliography}{9}
\bibitem{Sahni}  Varum Sahni, \textit{Lec. Notes Phys}. \textbf{653}, 141
(2004). ArXiv: astro-ph/0403324v3.

\bibitem{Faraoni}  T. P. Sotiriou and V. Faraoni, \textit{Rev. Mod. Phys}. 
\textit{\ }\textbf{82}, 451 (2010).

\bibitem{Santosa}  E. Santos, \textit{Phys. Rev. }\textbf{D81}, 064030
(2010).

\bibitem{Stabile}  A. Stabile, \textit{Phys. Rev. }\textbf{D82}, 124026
(2010); S. Capozziello and A. Stabile, arXiv: gr-qc/1009.3441v1 (2010).

\bibitem{Birrell}  N. D. Birrell and P. C. W. Davies, \textit{Quantum fields
in curved space. }Cambridge University Press, 1982\textit{.}

\bibitem{Wald}  R. M. Wald, \textit{Phys. Rev. }\textbf{D} \textbf{17}, 1477
(1978); \textit{Quantum field theory in curved spacetimes and black hole
thermodynamics, }The university of Chicago Press, Chicago, 1994.

\bibitem{Fischbach}  E. Fischbach and C. L. Talmadge, \textit{Nature} 
\textbf{356}, 207 (1992); ArXiv hep/th/96052249.
\end{thebibliography}
\end{document}